\documentclass[12pt]{article}
\pdfoutput = 1
\usepackage{cite}
\usepackage{graphicx}
\usepackage{amssymb}
\usepackage{amsthm}
\usepackage{amsfonts,amsmath}
\usepackage{appendix}
\usepackage{tikz} 
\usetikzlibrary{fit,positioning}
\usepackage[section]{placeins}
\usepackage[none]{hyphenat} 
\usepackage{lineno}
\usepackage{caption}
\usepackage{subcaption}
\usepackage{float}
\usepackage{hyperref}
\usepackage{colortbl}
\tikzstyle{int}=[draw, fill=white!8, minimum size=2em]
\tikzstyle{init} = [pin edge={to-,thin,black}]

\usepackage{tikz}
\usetikzlibrary{matrix}
\usetikzlibrary{calc}
\usetikzlibrary{arrows}
\usepackage{relsize}
\usepackage[export]{adjustbox}
\usepackage{array,longtable,multirow}

\usepackage{xcolor}

\title{Early pathogen replacement in a model of Influenza and Respiratory Syncytial Virus with partial vaccination. A computational study}
\author{Yury E. Garc\'ia, Marcos A. Capistr\'an \footnote{Centro de Investigaci\'on en Matem\'aticas A.C. Jalisco S/N Col. Valenciana, CP: 36240, Guanajuato, Gto. M\'exico. E-mail addresses: yury@cimat.mx (Y.E. Garc\'ia), marcos@cimat.mx (M.A. Capistr\'an)}}

\begin{document}
\date{\today}
\maketitle
\sloppy 

\begin{abstract}
In this paper, we carry out a computational study using the spectral decomposition of the fluctuations of a two-pathogen epidemic model around its deterministic attractor, i.e., steady state or limit cycle, to examine the role of partial vaccination and between-host pathogen interaction on early pathogen replacement during seasonal epidemics of influenza and respiratory syncytial virus.
\end{abstract}

\noindent%
\vfill
\section{Introduction}\label{sec:introduction}

In this paper, we study the impact of partial vaccination and between-host pathogen interaction on early pathogen replacement in a two-pathogen epidemic model of influenza and respiratory syncytial virus (RSV).\\

It is known that influenza and RSV peak during the winter season in temperate regions, and have semi-annual activity near tropical areas \cite{bloom2013latitudinal}. Second, the outbreak interference between these viruses have been registered by years  \cite{anestad1982interference,aanestad1987surveillance,anestad2009interference,glezen1980influenza,shinjoh2000vitro, welliver2008respiratory}. Influenza is known to interact with other viruses, including RSV. Glenzen et al \cite{glezen1980influenza} studied the interaction between influenza and other respiratory viruses; one of their conclusions is that simultaneous viral infections are a competition for resources and the virus with the largest growth rate is the one that succeeds in the invasion. There are also in vitro experiments presented by Shinjoh et al \cite{shinjoh2000vitro} who have shown that the growth of RSV can be blocked by influenza A if they infect the host cells at the same time. In contrast, RSV can suppress the growth of influenza A if this occurs after RSV infection.  Third, the effect of vaccination on epidemic synchrony patterns has been analyzed for several diseases.  In particular, Rohani et al. \cite{rohani1999opposite} have shown that vaccination turned synchronous epidemics (measles) into irregular and spatially uncorrelated epidemics once vaccination was deployed,  while whooping cough shifted from incoherence and spatial irregularity to regular dynamics as vaccination was introduced. Furthermore, in the context of vaccine-induced strain replacement, Martcheva et al \cite{martcheva2008vaccine} claim that {\it ``...the deployment of vaccination changes the proportion of hosts susceptible to either strain, ultimately shifting their relative and absolute abundances..."}. On the other hand, Alonso et al \cite{alonso2007stochastic} explain transitions in epidemics between regular and irregular dynamics in terms of amplification of demographic noise.\\

Based on the above results, in this paper, we have proposed a two-pathogen epidemic model with seasonality 
and partial vaccination as a continuous-time Markov jump process  using typical kinetic parameter values of 
influenza and RSV. Furthermore, we have used standard theoretical and computational methods \cite{van2002reproduction,kamo2002effect,adams2007influence,alonso2007stochastic,black2010stochastic}
to show that variations in coverage and efficacy of influenza vaccination may explain early pathogen 
replacement, e.g. either pathogen might invade first, and there is a second wave of infections where the 
second pathogen is dominant, see Anestad \cite{anestad1982interference,aanestad1987surveillance}.\\

The joint probability distribution of the state variables in our model is governed by a \textit{forward Kolmogorov} equation \cite{gardiner1986handbook}. The van Kampen asymptotic expansion \cite{van1992stochastic}  applied to this equation separates state variables into a mean field equation that matches the thermodynamic limit of the stochastic process (and is amenable to stability analysis in the sense of van den Driessche and Watmough \cite{van2002reproduction}), and a Fokker-Plank equation governing the system fluctuations, which is equivalent to a Langevin equation  in a neighborhood of the system attractor, e.g., steady state or limit cycle  \cite{alonso2007stochastic,black2010stochastic}. \\

First, we have used the mean field equation to carry out a standard analysis of the disease-free equilibrium
in terms of the effective vaccination rate and cross-immunity parameter.\\

Next, we have applied the Mckane approximation \cite{mckane2005predator} of the power spectral density  
(PSD) of the system fluctuations in a neighborhood of the system attractors \cite{alonso2007stochastic,rozhnova2009fluctuations,black2010stochastic, rozhnova2010stochastic,rozhnova2011stochastic}. Of note, this method of McKane to approximate the spectral decomposition of the system fluctuations is well suited to examine the role of between-host pathogen interaction and partial vaccination in seasonal patterns of respiratory diseases beyond the qualitative analysis of the mean field equation. Although, care must be taken since the PSD approximation does not hold near bifurcation points. \\

Coexistence of stable attractors in one and two pathogen epidemic models with seasonality has been documented \cite{kuznetsov1994bifurcation,kamo2002effect} at high contact rates.  Likewise, it has 
been established that the intertwined basin of multiple attractors at low contact rate,  or high vaccination 
the rate is not robust to spatial coupling \cite{alonso2007stochastic}. Since in this paper,  we care about the role of partial vaccination in seasonal epidemics, we shall focus our analysis in the yearly regime in a 
two-pathogen epidemic model with low spatial coupling aiming at showing a route to pathogen 
switching in early epidemic season.\\

The paper is organized as follows.  Section \ref{sec:MatMet} describes the mathematical model and the power spectral density for both seasonally forced and unforced models. Section 
\ref{sec:Results} shows the results when some key epidemic parameters are varied. Finally,  Section~\ref{sec:discussion} discusses our findings and offer some perspectives.

\section{Theoretical background}
\label{sec:MatMet}

\subsection{A Two-pathogen Epidemic Model with Partial Vaccination}
\label{sec:model}

The nonlinear dynamics of infectious disease spread in communities is stochastic. Assuming spatial homogeneity, the populations of susceptible, infectious and recovered individuals follow a birth and death process in $\mathbb{N}^{n}$.  Consequently, the epidemic model is posed as a continuous-time Markov jump process, whose forward Kolmogorov  equation is known as the Chemical Master Equation (CME). 

\begin{figure}[th]
\centering
\begin{tikzpicture}[node distance=2.5cm,auto,>=latex']
    \node [int] (SS) {$X_{SS}$};
    \node [int, fill=blue!5!white] (SI) [right of=SS] {$X_{SI}$};
    \node [int, fill=blue!5!white] (SR) [right of=SI] {$X_{SR}$};
    \node [int, fill=blue!5!white] (IS) [below of=SS] {$X_{IS}$};
    \node [int, fill=blue!5!white] (RS) [below of=IS] {$X_{RS}$};
    \node [int, fill=blue!5!white] (IR) [below of=SR] {$X_{IR}$};
    \node [int, fill=blue!5!white] (RR) [below of=IR] {$X_{RR}$};
    \node [int, fill=blue!5!white] (RI) [right of=RS] {$X_{RI}$};
    \node [int, fill=blue!5!white] (II) [below of=SI] {$X_{II}$};
    \node [int, fill=blue!5!white, pin={[init]above left:$\mu$}](SS) {$X_{SS}$}; 
    \node [coordinate] (end) [right of=SR, node distance=2cm]{};

    \draw[->] (SS) -- (1.0,-1.0) [above right] node {$\mu$};
    \draw[<-] (SI) -- (2.5,1.0) [above right] node {$\eta$};    
    \draw[->] (SI) -- (3.5,-1.0) [above right] node {$\mu$};
    \draw[->] (SR) -- (6.0,-1.0) [above right] node {$\mu$};  
    \draw[->] (IS) -- (1.0,-3.5) [above right] node {$\mu$};
    \draw[<-] (IS) -- (-1.0,-2.5) [below left] node {$\eta$};    
    \draw[->] (RS) -- (1.0,-6.0) [above right] node {$\mu$};
    \draw[->] (RI) -- (3.5,-6.0) [above right] node {$\mu$};
    \draw[->] (RR) -- (6.0,-6.0) [above right] node {$\mu$};
    \draw[->] (IR) -- (6.0,-3.5) [above right] node {$\mu$}; 
    \draw[->] (II) -- (3.5,-3.5) [above right] node {$\mu$};
    \draw[->] (0.5,1.8)  --(4.5,1.8) node [above,midway] {RSV}; 
    \draw[->] (-2.5,-0.5)--(-2.5,-5.0) node[above, midway,rotate=90] {Influenza};
       
    \path[->] (SS) edge node {$\beta_2 \lambda_2$} (SI);
    \path[->] (SI) edge node {$\gamma$} (SR);
    \path[->] (SR) edge node {$\sigma \beta_1 \lambda_1$} (IR);
    \path[->] (SS) edge [left]  node {$\beta_1 \lambda_1$} (IS);
    \path[->] (IS) edge [left]  node {$\gamma$} (RS);
    \path[->] (RS) edge node {$\sigma \beta_2 \lambda_2$} (RI);
    \path[->] (RI) edge node {$\gamma$} (RR);
    \path[->] (IR) edge node {$\gamma$} (RR);
    \path[->] (SS) edge [loop left,  looseness=1] node {$\upsilon$} (RS);   
    \path[->] (IS) edge node {$\sigma \beta_2 \lambda_2$} (II);
    \path[->] (II) edge node {$\gamma$} (IR);
    \path[->] (SI) edge [left]  node {$\sigma\beta_1 \lambda_1$} (II);
    \path[->] (II) edge [left]  node {$\gamma$} (RI);
\end{tikzpicture}
\caption{SIR model with two pathogens.  The first subscript denotes the infection status of 
influenza and the seconds subscripts denotes the infection status of RSV.  Labels by the 
arrows represent the reaction rates for each reaction type. Parameter definitions and 
dimensions are summarized in Table~\ref{tab:pars}.} 
\label{fig:model1}
\end{figure}
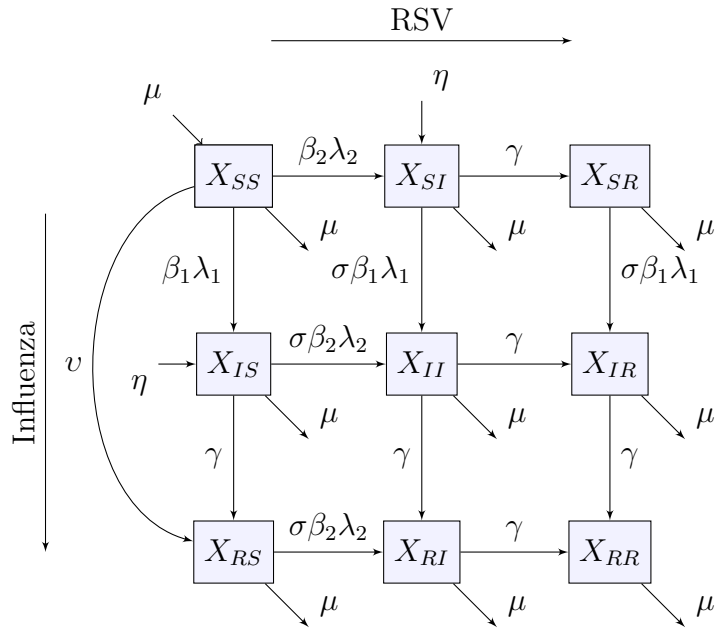

To model the dynamics of two pathogens with partial vaccination, we extend a SIR model 
following \cite{alonso2007stochastic} and \cite{kamo2002effect}. Let $X_{kl}(t)$  denote the 
number of individuals at time $t$ in immunological status $k\in\{S , I, R\}$ for pathogen 1 
(Influenza) and immunological status $l\in\{S, I, R\}$ for pathogen 2 (RSV). Reactions are 
illustrated in Figure \ref{fig:model1}.\\

We use mass action with contact rates $\beta_1$ and $\beta_2$ to describe the flow of  newly infected individuals from the susceptible group to the group of individuals infected with 
influenza or RSV respectively. The low spatial coupling is modeled with the immigration of infectious individuals with either disease at rate $\eta$. The average residence time for both diseases is  $1/\gamma = 7$ days \cite{CDC2017}. The population is assumed constant $\Omega$.  Therefore, we have set the birth rate equal to $\mu \Omega$, while life expectancy is set equal to $1/\mu=70$ years \cite{WHO2017}. Vaccination is not completely effective. Thus $\upsilon$ represents the effective vaccination rate. Vaccinated people either go to the recovered class $X_{RS}$  (concerning influenza) or remain in the susceptible class $X_{SS}$ depending on the vaccine efficacy. On the other hand, $\lambda_1$ and  $\lambda_2$ represent the population infected with influenza and RSV respectively.  Finally, to describe the relationship between RVS and influenza, we use a parameter $\sigma$ to describe either cross-immunity or cross-enhancement \cite{kamo2002effect,adams2007influence}. There is pathogen  cross-immunity when $0< \sigma<1$. This indicates that the presence of either pathogen inhibits the presence of the other one. $\sigma=0$ confers complete protection against a secondary infection and  $\sigma=1$ confers no protection. While $\sigma>1$ represents increasing the degree of cross-enhancement, i.e., the presence of either pathogen enhances the presence of the other one~\cite{adams2007influence}. 

\begin{table}[h!]
\centering
\resizebox{1.\textwidth}{!}{\begin{minipage}{\textwidth}
\begin{tabular}{lp{3cm}cp{2cm}cp{2cm}cp{2cm}}
Name  & Symbol & Value &Dimension \\
 \hline
Baseline contact rate  & $\beta_{i}$, $i=1,2$ & &$year^{-1}$ \\
Fraction of infectious individuals  & $\lambda_{i}$, $i=1,2$ & &1 \\
Cross immunity coefficient &  $\sigma$ & $[0,2]$&  $1$ \\
Effective vaccination rate & $\upsilon$ & $[0,1]$ & $year^{-1}$  \\
Immigration rate & $\eta$ & $12$ & $year^{-1}$ \\
Death/birth rate  &  $\mu$ &  $1/70$ & $year^{-1}$ \\ 
Recovery rate & $\gamma$ & $52.14$ & $year^{-1}$\\
 \hline
\end{tabular}
\caption{Two pathogen model parameters. Here $\beta_{i}$, for $i=1,2$ is the 
contact rate for influenza and RSV respectively.}
\label{tab:pars}
\end{minipage} }
\end{table}

\subsubsection{Chemical Master Equation}

Let us consider a closed population of size $\Omega$ at a given time t,  well mixed and homogeneously 
distributed, where individuals interact via $\mathcal{R}=25$ reactions depicted in Figure~\ref{fig:model1}. 
Transitions between states depend only on the time interval but not on absolute time, i.e.,  $X(\Delta t)$ 
and  $X(t+\Delta t)-X(t)$ are identically distributed. Additionally, two or more transitions take place in the same time interval with zero probability. Finally, for small time increments $\Delta t$, the transition 
probabilities $a_j(y)$ are obtained by multiplying the rates shown in Figure \ref{fig:model1} by $\Delta t$,
see~\cite{allen2008introduction,gillespie2007stochastic}. These assumptions are encoded in the Kolmogorov 
forward equation (Chemical Master Equation, or CME). It represents the evolution of the probability distribution of finding the system in state $X=x$ at time t.  

\begin{equation}\label{eqs:ME}
\dfrac{dP_x(t)}{dt} = \sum_{j=1}^{\cal R}a_j(x-v_j)P_{x-v_j}(t)-\sum_{j=1}^{\cal R}a_j(x)P_x(t)
\end{equation}

where $x(t)$ corresponds to the realizations of the random vector $X(t)=[X_{i}(t)]$ and $v_j(t)$ are the 
stoichiometric vectors e.g. vectors whose elements in $\{-1, 0 ,1\}$ describe the addition/subtraction of 
mass from a particular compartment.  Let $S = S_{ij},  i=1,...,5$; $j=1,...,\mathcal{R}$ be the stoichiometric matrix that describes changes in the population size due to each of the $\mathcal{R}$ reactions and  $S=[v_1,\ldots,v_{\mathcal{R}}]$.  A list with the $\cal R$ reactions and the explicit form of these terms are defined in the supplementary material.

\subsubsection{Seasonal Forcing}

Often, in order to analyze the full time-dependent master equation for the two pathogens model with seasonal 
forcing, authors describe the system dynamics using the same equations, e.g. equations (\ref{eqs:ME}) and 
(\ref{eq:Langevin}), except that $\beta_1$ and $\beta_2$ are functions of time, i.e.,

\begin{equation}\label{eqs:beta}
\beta_{p}(t) = \beta_{i}(1+\delta \cos (2\pi t/T))
\end{equation}

for $p,i=1,2$. Parameters $\beta_{i}$,  are the baseline contact rate, $\delta$ is the magnitude of seasonal 
forcing and $T$ is the period of one year.  

\subsection{Theoretical and computational tools}
\subsubsection{Van Kampen Expansion}

For large populations, equation (\ref{eqs:ME}) is computationally too expensive 
to be solved exactly. Hence, we assume that the linear noise approximation holds

\begin{equation}\label{eqs:LNA}
X(t) = \Omega \phi(t) + \Omega^{1/2} \xi(t), \quad t\in[0, T],
\end{equation}

namely,  for large $\Omega$ the system states $X=x$ can be expressed as the sum of a 
macroscopic term $\phi(t)$ and a stochastic term $\xi(t)$, which describes the fluctuations and accounts for demographic stochasticity in the system. Combining equations~(\ref{eqs:ME}) and~(\ref{eqs:LNA}) 
gives rise to the van Kampen expansion \cite{van1992stochastic}. Assuming constant average concentration, the size of the stochastic component will increase as the square root of population size. 
The time-evolution of the terms of order $\Omega^{1/2}$ \cite{van1992stochastic} is governed by the ODE 
system

\begin{equation}
\label{eqs: detpartODE}
\begin{split}
\dfrac{d \phi_i (t)}{dt} & =\sum_{j=1}^{\cal R} S_{ij}f_j(\phi(t),t)\\
\phi_i (0) &= \phi_0
\end{split}
\end{equation}

where  $t\in [0,T], \; i=1,\ldots, \mbox{dim}\{X(t)\}$,  
$\phi_i(t)=\lim_{\Omega, X \longrightarrow \infty}X_{i}/\Omega$, and $f_j(\phi(t),t) =  a_j(\phi(t))$. Collecting terms of order $\Omega^0$, we obtain a Fokker Plank equation for the joint distribution of the system fluctuations, see \cite{van1992stochastic}. Of note, there is a well known Langevin equation, which describes the temporal evolution of the normalized fluctuation of susceptible and infectious 
states \cite{alonso2007stochastic}, and whose solution is the same as the Fokker-Planck
equation for the system fluctuations in a neighborhood of the macroscopic steady state

\begin{equation}\label{eq:Langevin}
\dot{\xi}(t) = A(t)\xi(t)+\zeta(t)
\end{equation}

where $\zeta(t)$ is white noise with zero mean and correlation structure given by  $\langle\zeta(t)
\zeta(t')^T\rangle = B(t)\delta(t-t')$. Here, $A(t) = \partial S f(\phi(t),t)/\partial \phi(t)$,
$B(t) = EE^{T}$ and $E = S\,\mbox{diag}\{\sqrt{f(\phi(t), t)}\} $, see 
\cite{van1992stochastic,gillespie2007stochastic,komorowski2009bayesian}. 

\subsubsection{Power Spectral Density}
We consider both, seasonally forced and unforced models. Our contributions rest on examining how the natural frequency of the epidemic outbreak varies when some key epidemic parameters are changed \cite{wang2012simple}. Consequently, in this Subsection, we describe the method first introduced by Newman and Mckane \cite{reuman2006power} to compute the analytical PSD in a neighborhood of the system attractor (steady state or limit cycle) to a two pathogen model. 

\subsubsection{Unforced Model}
\label{subsec:unforced}

We consider the power spectral density of the fluctuations obtained through Wiener-Khinchin theorem, \cite{champeney1987handbook} by Fourier transforming 
linear stochastic differential equation (\ref{eq:Langevin})

\begin{displaymath}
P_k(\omega) = \langle|\tilde{\xi}_k(\omega)|^2\rangle
\end{displaymath}

formally 

\begin{displaymath}
\tilde{\xi}_k = \int_{-\infty}^{\infty}\xi_k(t)e^{-k\omega t}dt
\end{displaymath}

for $k = 1,\ldots,5$. Rozhnova~\cite{rozhnova2011stochastic}, \cite{rozhnova2009fluctuations} provides 
a closed expression for the PSD in terms of matrices $A$ and $B$ obtained by the van Kampen expansion. 
To compute the PSD, matrices $A$ and $B$ are evaluated at the steady state of the system (\ref{eqs: detpartODE}).  The general solution is given by 

\begin{equation}
\label{eq:psd}
P_{kl}(\omega)  = \sum_{i,j} \Phi_{kj}(\omega)B_{ji} \Phi_{il}^\dagger(\omega), 
\end{equation}

$\Phi^\dagger(\omega) = (\Phi^H)^{-1} $ means the inverse of the conjuate transpose of $\Phi$ where $\Phi(\omega)=-i\omega I-A$. Equation~(\ref{eq:psd}) allows to compute the PSD for a wide 
range of frequencies and parameter ranges with moderate computational burden.

\subsubsection{Forced Model}
\label{subsec:Floquet}

Matrices $A(t)=A(t+T)$ and $B(t)=B(t+T)$ are now periodic functions of time, instead of the method used in Subsection~\ref{subsec:unforced}, we use Floquet's theory to find the solution of Eqs. (\ref{eq:Langevin}) 
and compute its power sprectrum density, see~\cite{black2010stochastic}. \\

The solution of equation (\ref{eq:Langevin}) can be written as a sum of the general solution of the 
homogeneous and a particular solution of the inhomogeneous system getting
\begin{equation}
\frac{d\Phi(t)}{dt} = A(t)\Phi(t)
\end{equation}

Where $\Phi(t)$ is the fundamental matrix \cite{grimshaw1991nonlinear}, formed from the linearly 
independent solutions of  homogeneous equation $\dot{\xi}(t) = A(t)\xi(t)$. Floquet's theorem states 
that there exists a periodic non singlular matrix  $M$ \cite{grimshaw1991nonlinear} such that 
$$\Phi(t+T)=\Phi(t)M$$

Matrix $M$ is sometimes referred as the monodromy matrix of the 
fundamental matrix $\Phi(t)$. This can be expressed in terms of the fundamental matrix by setting $t=0$
\begin{equation}
M = \Phi^{-1}(0)\Phi(T)
\end{equation}

It is useful to choose $\Phi(t)$ to be the principal matrix, so that $\Phi(0)=I$, and then $M=\Phi(T)$. The eigenvalues of $M$, $\rho_1, \ldots, \rho_n$, are called the \textit{caracteristic multipliers} and a related set of quantities are the \textit{Floquet exponents} defined by
\begin{equation}\label{FloExp}
\vartheta_i = \dfrac{ln(\rho_i)}{T}
\end{equation}

Of note, a limit cycle will be stable if $|\vartheta_i|<1$, see  \cite{rozhnova2010stochastic}, \cite{grimshaw1991nonlinear}. Using further Floquet's theory and analytical expression, it is possible to obtain the auto-correlation function of the stochastic fluctuations \cite{black2010stochastic}, \cite{boland2009limit}, \cite{rozhnova2010stochastic}, given by

\begin{equation}
C(\tau) = \dfrac{1}{T}\int_0^{T}\langle \xi(t+\tau)\xi^{'}(t)dt\rangle, \hspace{0.3cm}  \xi \equiv \{\xi_1,\ldots,\xi_5\}
\end{equation}

Taking the Fourier transform of this expression, we get an exact expression for power spectrum 
of the stochastic oscillations. The details are presented in references \cite{black2010stochastic, boland2009limit}  and the algorithm to compute the PSD is described by Black \cite{black2010thesis} (p. 111).

\subsubsection{Coherence}

Based on Alonso et al \cite{alonso2007stochastic} definition of coherence as a measure of stochastic 
amplification, we consider the normalized cross-correlation as a measure of similarity of influenza 
and RSV spectral densities

\begin{equation}
\label{eq:coherence}
Q_{kl}(\omega)=\frac{P_{kl}(\omega)}{\sqrt{P_{k}(\omega)P_{l}(\omega)}},
\end{equation}

where we denote $P_{kl}(\omega)=P_{k}(\omega)$ if $k=l$. In Section \ref{sec:Results} we 
use equation \eqref{eq:coherence} to examine the out of phase relationship and correlation of influenza 
and RSV signals as a function of vaccination and cross-immunity rates at selected frequency ranges.

\section{Results}\label{sec:Results}

To carry out our analysis of the two pathogen model, we take a simplified Markov 
jump process whose elements are defined in terms of the Markov process 
defined in Figure \ref{fig:model1} using the following identities

\begin{align*}
Y_1(t) &= X_{SS}(t)\\
Y_2(t) &= X_{IS}(t)+X_{II}+X_{IR}(t)\\
Y_3(t) &= X_{SI}(t)+X_{II}+X_{RI}(t)\\
Y_4(t) &= X_{IS}(t)+X_{RS}(t)\\
Y_5(t) &= X_{SI}(t)+X_{SR}(t)
\end{align*}

where the $Y_{i}$, $i=1,...,5$ represent respectively, the number of those individuals who are susceptible 
to both pathogens, those who are infected with influenza only, those who are infected with RSV only, those 
who are susceptible to RSV and those who are susceptible to influenza respectively. 

\subsection{Role of seasonality on system fluctuations}

According to Rozhnova and Nunes \cite{rozhnova2010stochastic}, using Floquet's theory described in (\ref{subsec:Floquet}) we can show that the power spectral density has peaks at frequencies 

\begin{equation}\label{eqs:frecuency}
\dfrac{m}{T}\pm \dfrac{|Im(\vartheta_p)|}{2\pi}, \hspace{0.3cm} p=1,2
\end{equation}

where $m$ is an integer and $\vartheta_p$ are the Floquet exponents. For the annual limit-cycle 
the dominant peak is at $Im(\vartheta_p)/2\pi$, with the others peaks being much smaller. Here, 
$|Im(\vartheta_p)|$ denotes the absolute value of the imaginary part of complex conjugate Floquet 
exponents. \\

\begin{figure}[htbp]
\includegraphics[width=1.0\textwidth,center]{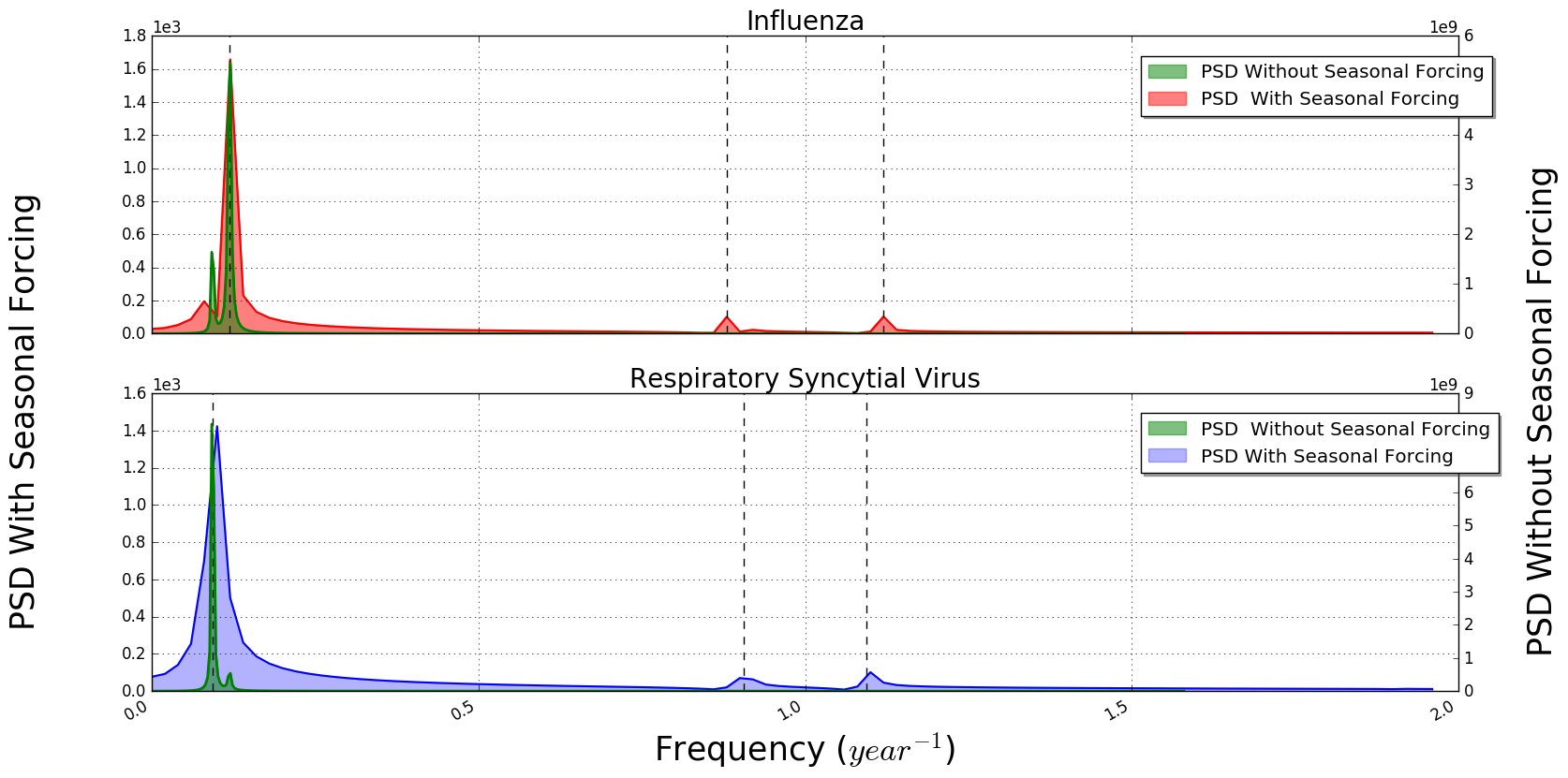}\\
\caption{{\bf PSD without and with seasonal forcing}. Vertical helper lines mark the frequencies predicted by Eq. (\ref{eqs:frecuency}). The parameters values are $\beta_1 = 93.88$ $year^{-1}$, $\beta_2 = 83.45$ $year^{-1}$, $\sigma=0.8$, $\delta=0.06$ and $\upsilon=0$. The Floquet exponents are $\vartheta_1 = -0.01341813\pm 0.74852756i $ and $\vartheta_2 = -0.01071943+0.58598082i $ and $\vartheta_3=-0.02753316+0.i$. } 
\label{fig:FigureSF1}
\end{figure}

Figure \ref{fig:FigureSF1} shows the  analytic PSD with and without seasonal forcing. The parameters 
values are $\beta_1 = 93.88$ $year^{-1}$, $\beta_2 = 83.45$ 
$year^{-1}$, $\sigma=0.8$, $\delta=0.06$ and $\upsilon=0$.  Floquet exponents are given by 
$\vartheta_1 = -0.01341813\pm 0.74852756i $ and $\vartheta_2 = -0.01071943+0.58598082i $ 
and $\vartheta_3=-0.02753316+0.i$. Thus, the dominant peak for Influenza is given by 
$Im(\vartheta_1)/2\pi= 0.119131$ $year^{-1}$ and the mean peak for RSV is given by 
$Im(\vartheta_2)/2\pi = 0.0932$ $year^{-1}$. We can see that the PSD for the non-seasonal case 
are qualitatively comparable with the PSD obtained for the seasonally forced system. The main 
difference is the addition of two annual peaks with seasonal transmission. The period for both peaks 
of influenza are $0.893549$ and $1.1352436$ $year$ and the period for the RSV peaks are $0.914694$ 
and $1.1028540$ $year$.\\

Bifurcation diagram (not shown), indicates that there is a period doubling bifurcation as we increase 
either $\sigma$, or $\delta$. However, in the one hand we care about pathogen replacement during the 
early season epidemics. And on the other hand, the PSD analytic formulas do not hold near bifurcation points.
Consequently, we limit our analysis to the yearly regime.

\subsection{Epidemic criticality conditions}

Since seasonal and epidemic fluctuations are separated as indicated in Figure \ref{fig:FigureSF1}, 
in the remainder we  focus on the analysis of the epidemics fluctuations. Let us analyze the steady 
states of the system without seasonality to understand the relationship between effective vaccination 
and cross-immunity. In terms of the deterministic model~(\ref{eqs: detpartODE}), the largest eigenvalue 
of the next generation matrix, or basic reproductive number is

\begin{equation}
R_0 = \max_{i=1,2} R_i
\end{equation}

where, $R_{1}$, $R_{2}$ given by

\begin{equation}
\begin{split}
R_1 &= \dfrac{\mu\beta_1}{(\gamma+\mu)(\mu+\upsilon)}\\
R_2 &= \dfrac{\beta_2(\mu+\sigma\upsilon)}{(\gamma+\mu)(\mu+\upsilon)}
\end{split}
\end{equation}

are the two only eigenvalues of the next generation matrix of van den Driessche and Watmough \cite{van2002reproduction} (see supplementary material for details). The two eigenvalues correspond 
to the reproduction numbers for each pathogen, $R_1$ for influenza and $R_2$ for RSV.  If $\eta=0$ system (\ref{eqs: detpartODE}) has a disease free equilibrium ($Y_{DFE}$)

\begin{equation}
\label{FixPoint1}
Y_{DFE} = [\mu/(\mu+\upsilon),0,0,\upsilon/(\mu+\upsilon),0].
\end{equation}

The disease-free equilibrium is stable if $R_{0}<1$ and unstable if $R_{0}>1$. There is a  steady state where only people infected with influenza is present, and similarly, there is a steady state where only people infected with RSV is present. Also, there is a steady state where both diseases coexist. We base this claim on Vasco et al \cite{vasco2007tracking} analysis of the case with no vaccination.

\subsection{Analysis of cross-immunity}

Let us study the disease-free equilibrium (\ref{FixPoint1}), and conditions under which it is possible to eradicate both diseases assuming that there is no seasonal forcing. We will focus on the cross-immunity $\sigma$ and the effective vaccination rate $\upsilon$. \\

The equilibrium point $Y_{DFE}$ is stable if $R_0<1$. Of note, $R_1$ and $R_2$ dependence 
on $\upsilon$ imply that by increasing the effective vaccination value we may reduce $R_0$ 
below 1, thus erradicating both diseases, even when vaccination is only against influenza.  
$R_0<1$ implies  $R_1<1$ and $R_2<1$. We explore the scenarios that may take place given the conditions of stability for the free-disease equilibrium, i.e. $R_1<R_2<1$ and $R_2<R_1<1$. Consequently, we will plot the PSD for selected vaccination values in each case. \\

If we consider the case when  $R_1<R_2<1$, the conditions to effective vaccination rate and 
cross-immunity are

\begin{itemize}\label{condiction1}
\item $\upsilon>\dfrac{\mu(\beta_2-(\gamma+\mu))}{(\gamma+\mu-\sigma\beta_2)}$
\item $\sigma<\dfrac{\gamma+\mu}{\beta_2}<1$
\end{itemize}

On the other hand, if we consider the case $R_2<R_1<1$ we have:
\begin{itemize}\label{condiction1}
\item $\upsilon>\dfrac{\mu(\beta_1-(\gamma+\mu))}{(\gamma+\mu)}$
\item $\sigma<\dfrac{\gamma+\mu}{\beta_2}<1$ 
\item $\beta_1>(\gamma+\mu)$
\end{itemize}

There are three cases to consider,

\begin{description}

\item[Case $\sigma<1$.] See Fig. \ref{fig:vaccination}. This condition assumes that getting sick from either virus confers some protection against the second one.  Let us denote $\mathcal{R}_1$ and $\mathcal{R}_2$ 
respectively the values of  $R_1$ and $R_2$ without the presence of effective vaccination, 
i.e. $\mathcal{R}_1 =\dfrac{\beta_1}{\gamma+\mu}$ and 
$\mathcal{R}_2 = \dfrac{\beta_2}{\gamma+\mu}$. \\

Furthermore, let us consider the starting point in parameter space where $\mathcal{R}_1= 1.6$ 
and $\mathcal{R}_2=1.23$. This implies $\beta_1 = 83.45$ $year^{-1}$  and  $\beta_2 = 64.15$ 
$year^{-1}$. Let us set $\sigma=0.8$. With these values, $\sigma$ and $\upsilon$ satisfy the  condition above. Thus, the effective vaccination rate values for vanish both viruses are $\upsilon=0.008571$ $year^{-1}$ for influenza and  $\upsilon = 0.2053999$ $year^{-1}$ for syncytial.  This means that when there is some kind of cross-immunity between these viruses, it is possible to control both, even when vaccine acts only against influenza.  On the other hand, it is possible to find vaccination values that, while decreasing the amplification of influenza, increase the amplification of RSV.  It is noteworthy that there is a vaccination value ($\upsilon\approx0.0073$) where the PSD peak for both diseases trade places. 
Of note, this phenomenon has been observed in real data, e.g., \cite{noyola2005contribution}. \\

\begin{figure}[H]
\includegraphics[width=1.0\textwidth,center]{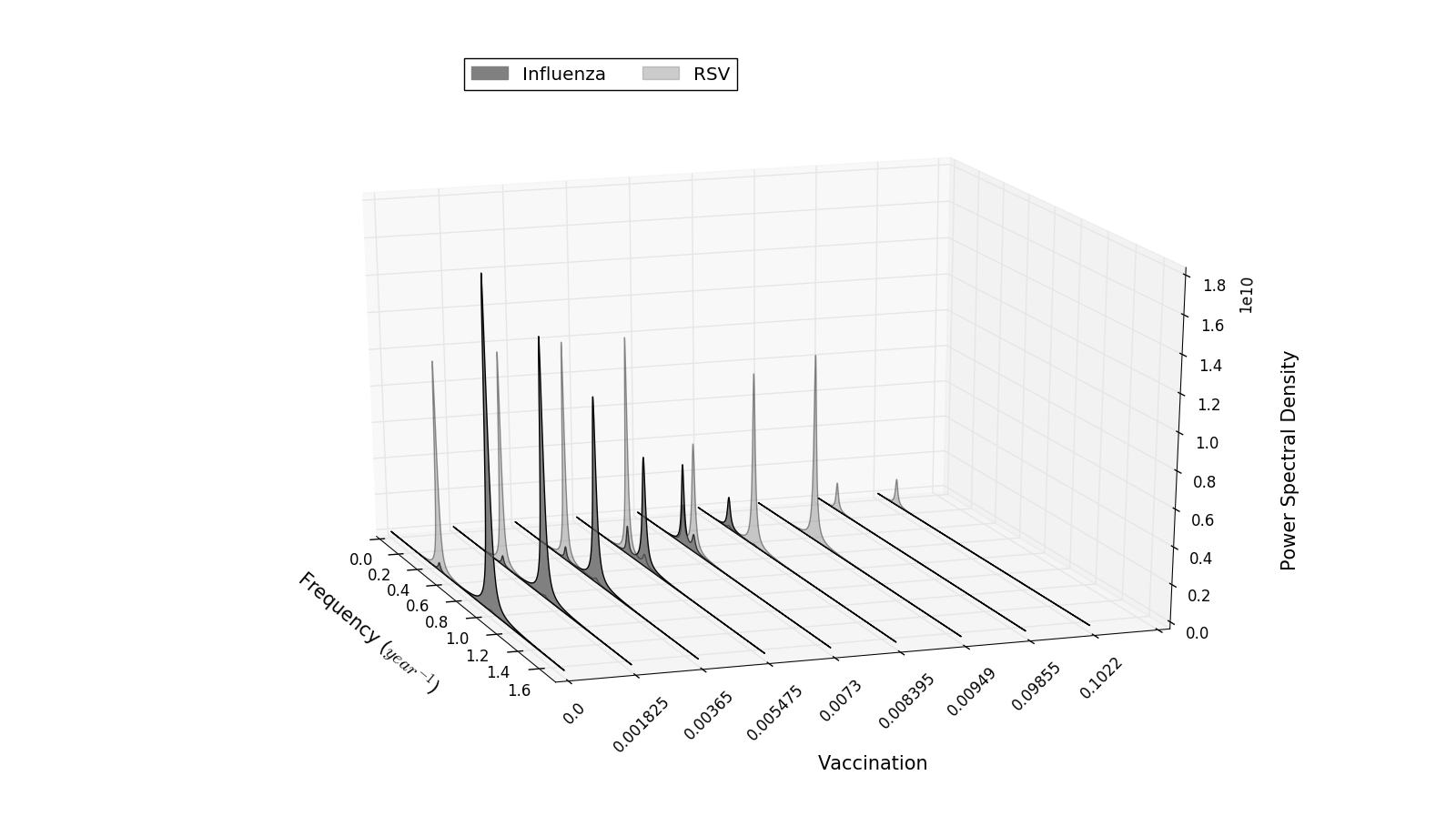}
\caption{Power Spectral Density for the case $\sigma<1$. This figure shows how the PSD for influenza and RSV vanish when crossing the respective thresholds $\upsilon=0.008571$ and $\upsilon = 0.2053999$ $year^{-1}$. PSD is shown at selected vaccination rates.}
\label{fig:vaccination}
\end{figure}

\item [Case $\sigma=1$] In this case the behavior of both pathogens is independent of each other. 
Note that $R_2 = \dfrac{\beta_2}{\gamma+\mu}$, i.e., it no longer depends on $ \upsilon $ or  $\sigma $.  Let us consider $\mathcal{R}_1=1.6$ and $\mathcal{R}_2=1.5$ implies $\beta_1 = 83.65$ and $\beta_2 = 78.42$\\

In this case, the behavior of both diseases is independent, see Fig. \ref{case2}. The frequency of the  RSV peaks is the same in all cases and the frequency of influenza peaks 
increases when the effective vaccination rate is increased. 

\begin{figure}[H]
\centering
\includegraphics[width=1.0\textwidth,center]{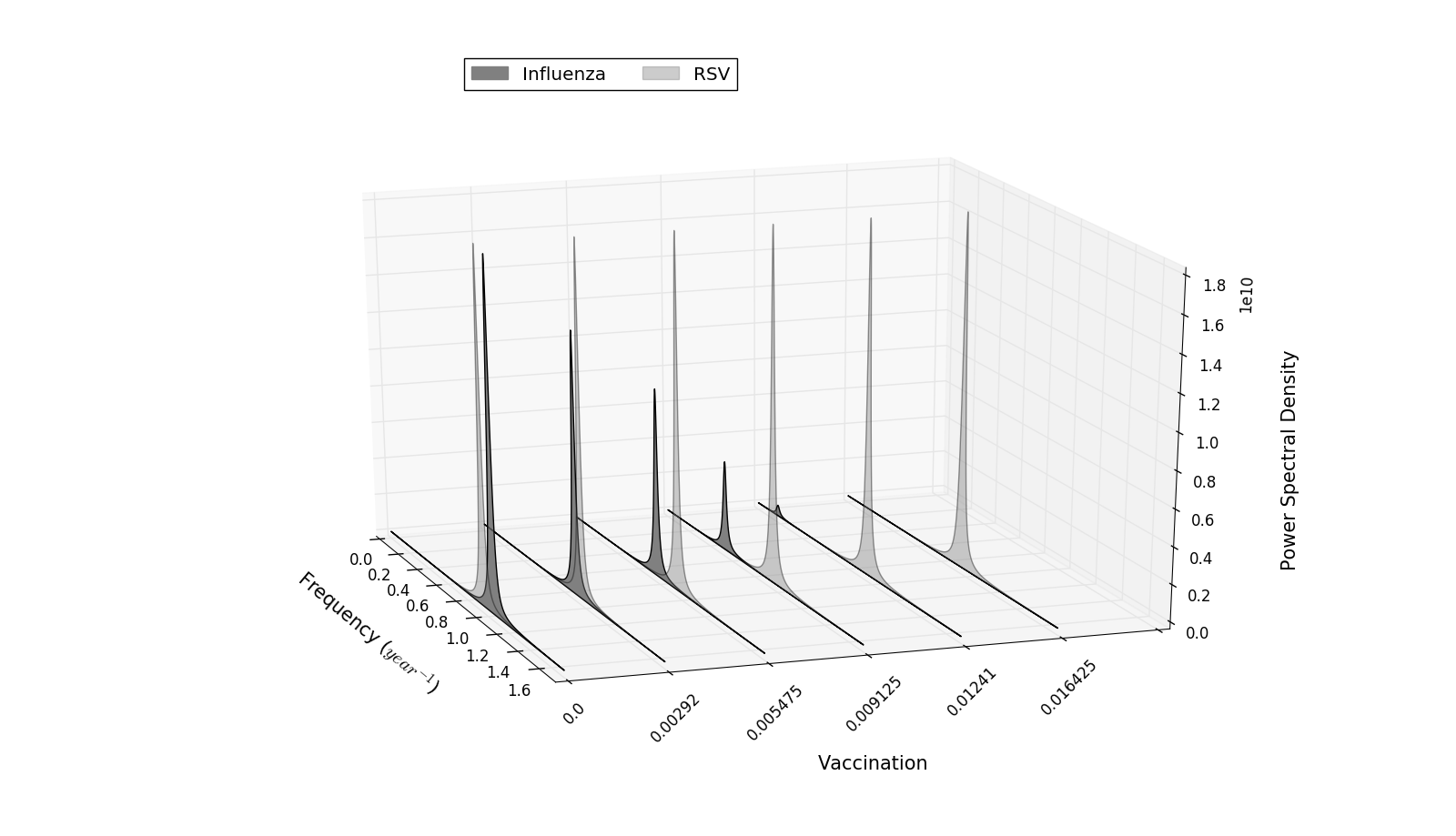}
\caption{Power Spectral Density for the case $\sigma=1$.  Parameter values are $\sigma=1.0$, $\beta_1 = 83.65$ and $\beta_2 = 78.42$.}
\label{case2}
\end{figure}

\item [Case $\sigma>1$.] 
Let us consider the case when $\sigma=1.1$, that means that the presence of one pathogen 
enhances the presence of the second one. Thus, the vaccination rate not only changes the 
period of the influenza peak but also produces small changes in the RSV peak period. \\

Figure \ref{fig:swap} provides evidence that vaccination affects both influenza and RSV total amplification.  Total amplification corresponds to the integral of the PSD over all frequencies.  This phenomenon is observed when $\sigma$ is less and greater than one.  After several experiments, we noticed that in some cases, by increasing the vaccination rate, we reduce the amplification of influenza but it can be amplified that of the RSV, see Fig. \ref{fig:vaccination} when $\upsilon \approx 0.0083$ and Fig. \ref{fig:swap} when $\upsilon \geq 0.0073$.

\begin{figure}[H]
\includegraphics[width=1.0\textwidth,center]{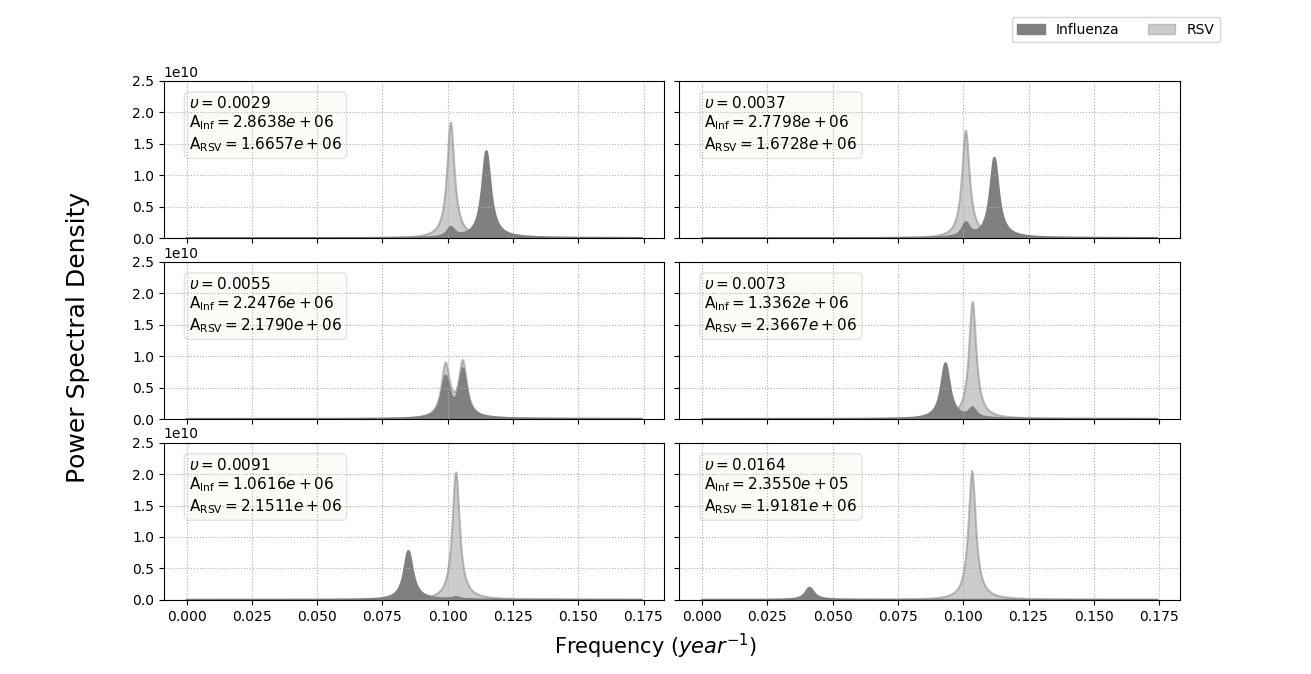}\\  %
\caption{Total Amplification for both pathogens. Parameter values: $\mathcal{R}_1=1.8$, $\beta_1=93.88$ $year^{-1}$, $\mathcal{R}_2=1.5$, $\beta_2=78.23$ $year^{-1}$, $\sigma=1.1$. $A_{Inf}$ and $A_{RSV}$ correspond to the total amplification of influenza and RSV respectively. They are calculated with different vaccination rate values ($\upsilon$)}
\label{fig:swap}
\end{figure} 
\end{description}

\section{Discussion}
\label{sec:discussion}

There is evidence that before the introduction of influenza vaccination programs, seasonal patterns of RSV and influenza were regular, with an outbreak of RSV immediately followed by an influenza outbreak each year\cite{aanestad1987surveillance}. But, in the presence of vaccination against influenza either pathogen might invade first, see \cite{rozhnova2010stochastic,noyola2005contribution}. Our analysis supports the claim that early season pathogen replacement depends on effective vaccination rate and relative virus fitness, e.g., $R_1$ and $R_2$.  Vaccination changes the natural frequency and relative fitness of both virus, thus allowing either virus to peak first in a given season. If the strength of influenza infection ($R_1$) is greater than the strength of the RSV infection ($R_2$) and the vaccination rate is small, then the influenza peak can happen first. Otherwise, if the vaccination rate is large enough the RSV  peak might happen first, even when $R_1>R_2$. However, the peak of RSV will appear first if $R_2>R_1$. \\

\begin{figure}[htbp]
\includegraphics[width=1.0\textwidth,center]{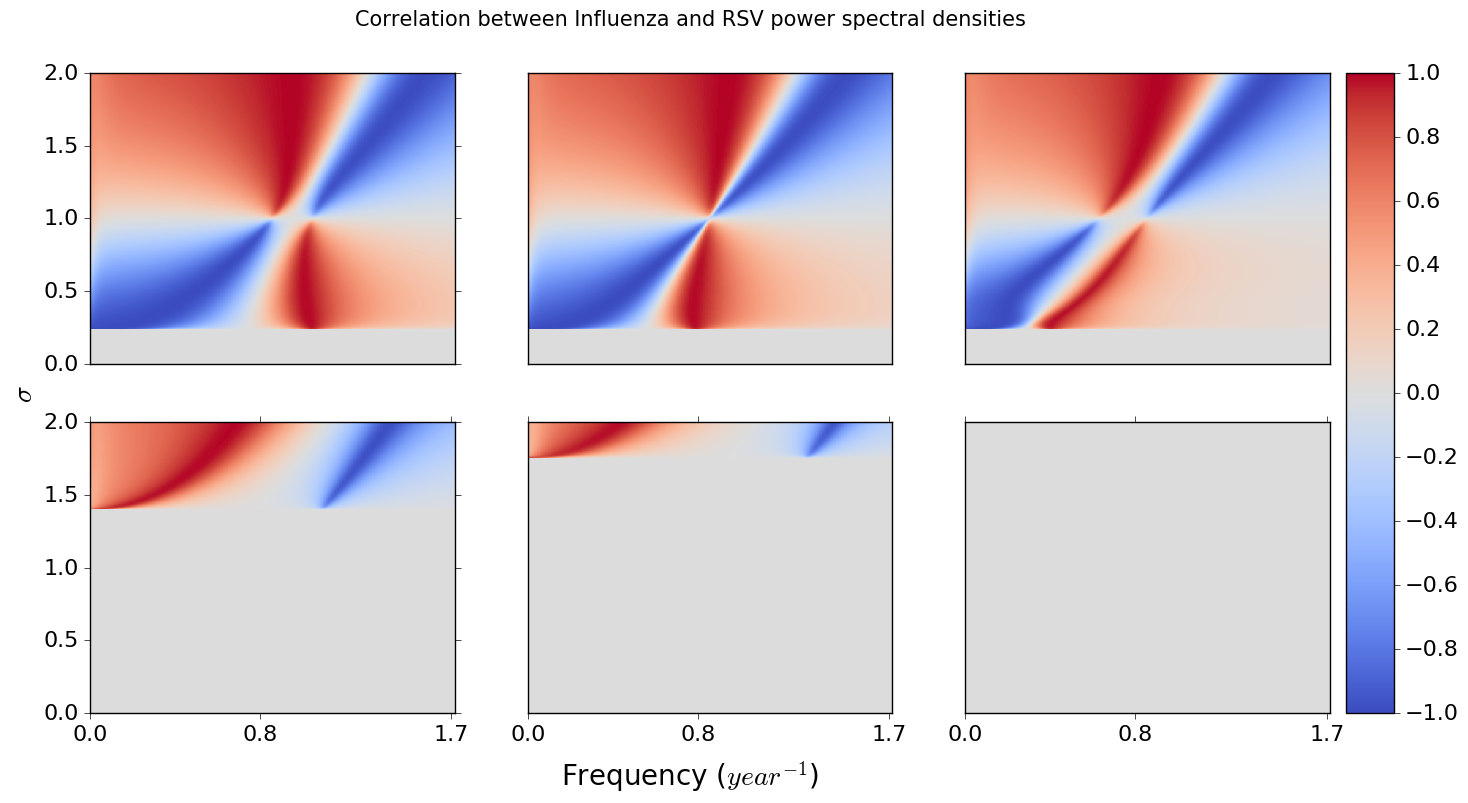}\\
\caption{{\bf PSD Correlation as a measure of stochastic amplification among pathogens.} Given fixed basic 
reproductive numbers for influenza ($\mathcal{R}_{1}=1.6$), and RSV ($\mathcal{R}_{2}=1.4$), we show from left to right, from top to bottom PSD Correlation at vaccination rates $\upsilon=(0,0.0036,0.0091,0.018,0.025,0.036)$ year$^{-1}$.  This figure shows the correlation between the two pathogen amplifications when $\sigma$ is varied and how the frequency for both viruses change when $\upsilon$  increase. For $\sigma =1$ they are completely independent.
}
\label{fig:correlation}
\end{figure}

Previous results \cite{anestad1982interference, noyola2005contribution, shinjoh2000vitro} support the existence of interference between outbreaks of RSV infection and influenza.  We include the term $\sigma$ in the model to explore the immunity relationship between pathogens at the population level. \\

When $\sigma\neq1$ there is a correlation in the fluctuations of both diseases as we can see in Figure \ref{fig:correlation}. Consequently, partial vaccination not only affects the behavior of influenza but also the behavior of RSV. Of note, according to the model, it becomes possible to eradicate RSV when $\sigma<1$ by increasing the rate of vaccination, even if vaccination is directed only against influenza.  On the other hand, $\sigma=1$ means that there is not immunity relation between the pathogens. In this case, 
we can vanish influenza by increasing the vaccination rate without having any effect on RSV.  Moreover, when $\sigma\approx1$  and $R_1\approx R_2$, the periodicity of both diseases are similar and there is an overlap giving the shape of M that we can see in real time series \cite{anestad1982interference}, \cite{noyola2005contribution}. But, when $R_1$ and $R_2$ have a considerable difference or $\sigma$ is far from one, the peaks have a totally different period. \\

Another important factor is the seasonal forcing. Influenza and RSV are seasonally related \cite{mangtani2006association}, \cite{bloom2013latitudinal}. Circulation of both often occur at similar times of the year in some temperate zones and peaks timing differ by less than one month\cite{bloom2013latitudinal}, \cite{velasco2015superinfection}. Seasonality can induce epidemic cycles. When this is included in the model, the peaks periodicity are not affected but, non-seasonal peaks appear in frequencies $m/T \pm |Im(\vartheta)|/2\pi$. We predict the number and position of the dominant and non-seasonal peaks as a function of the epidemiological parameters.  \\

Finally, we consider that our analysis might serve as a basis to explore further the effect of partial vaccination on multi-pathogen epidemics. Of particular importance is to study the effect of vaccination aiming at reducing the morbidity caused by respiratory diseases.

\bibliography{Bibliography}{}
\bibliographystyle{plain}
\end{document}


\date{\today}
\maketitle
\sloppy 

In this supplementary material we describe the complete reactions and the matrices used to obtain the master equation and the expression for the power spectral density for a two-pathogen system. 

\section{Model}

Equations \ref{eq:ODE1} correspond to the mean field of the system presented in Fig. 1 in the main paper.

\begin{equation}
\begin{split}
\dot{X_{SS}}(t) &= \mu\Omega - \beta_2\lambda_2X_{SS}-\beta_1\lambda_1X_{SS}-\mu X_{SS}-\upsilon X_{SS}\\
\dot{X_{IS}}(t) &= \beta_1\lambda_1 X_{SS}-\gamma X_{IS}-\mu X_{IS}-\sigma\beta_2\lambda_2 X_{IS}\\
\dot{X_{RS}}(t) &= \gamma X_{IS}-\sigma\beta_2\lambda_2 X_{RS}-\mu X_{RS}+\upsilon X_{SS}\\
\dot{X_{SI}}(t) &= \beta_2\lambda_2 X_{SS}-\gamma X_{SI}-\mu X_{SI}-\sigma\beta_1\lambda_1 X_{SI}\\
\dot{X_{RI}}(t) &= \sigma\beta_2\lambda_2 X_{RS}-(\gamma+\mu)X_{RI}+\gamma X_{II}\\
\dot{X_{SR}}(t) &= \gamma X_{SI}-\mu X_{SR}-\sigma \beta_1\lambda_1 X_{SR}\\
\dot{X_{IR}}(t) &= \sigma \beta_1 \lambda_1 X_{SR}-(\gamma+\mu) X_{IR}+\gamma X_{II}\\
\dot{X_{RR}}(t) &= \gamma X_{SR} + \gamma X_{RI}-\mu X_{RR}\\
\dot{X_{II}}(t) &= \sigma \beta_1\lambda_1 X_{SI}+\sigma\beta_2\lambda_2X_{IS}-(\mu+2\gamma)X_{II}
\end{split}
\label{eq:ODE1}
\end{equation}

where $\lambda_1 = (X_{IS}+X_{II}+X_{IR})/\Omega$ is the proportion of infected people with virus one (influenza), and $\lambda_2 = (X_{SI}+X_{II}+X_{RI})/\Omega$ is the proportion of infected people with virus two (RSV).  The meaning of the parameters and variables are described in the main document.  To simplify the calculation we  take a new Markov jump process $Y(t)$ whose elements are defined by the Markov process $X(t)$  above \cite{kamo2002effect, adams2007influence} using the following identities.
\begin{align*}
Y_1(t) &= X_{SS}(t)\\
Y_2(t) &= X_{IS}(t)+X_{II}+X_{IR}(t)\\
Y_3(t) &= X_{SI}(t)+X_{II}+X_{RI}(t)\\
Y_4(t) &= X_{IS}(t)+X_{RS}(t)\\
Y_5(t) &= X_{SI}(t)+X_{SR}(t)
\end{align*}

with this simplification we get a new system of differential equations from Eqs. (\ref{eq:ODE1}) given by
\begin{equation}\label{eq:ODE2}
\begin{split}
\dot{y_1}(t) &= -\beta_2y_3 y_1-\beta_1 y_2 y_1-\upsilon y_1+\mu(\Omega-y_1)\\
\dot{y_2}(t) &= \beta_1 y_2(y_1+\sigma y_5)-(\gamma+\mu)y_2\\
\dot{y_3}(t) &= \beta_2 y_3(y_1+\sigma y_4)-(\gamma+\mu)y_3\\
\dot{y_4}(t) &= \upsilon y_1 + \beta_1 y_2 y_1-\sigma\beta_2 y_3 y_4-\mu y_4\\
\dot{y_5}(t) &= \beta_2 y_3y1-\beta_1\sigma y_2 y_5-\mu y_5\\
\end{split}
\end{equation}

$y(t)=[y_1(t), \ldots, y_5(t)]$ correspond to the realization of the stochastic vector $Y(t)=[Y_1,\ldots,Y_5(t)]$. To model the presence of stochasticity, we translate the system of ordinary differential equations definded above (\ref{eq:ODE2}) into a stochastic process model. We do this by considering each flux between compartments to be random. Possible reactions are listed below:

\begin{table}[H]
\begin{center}
\begin{tabular}{lll}
\textbf{Reactions}     & \textbf{Propensity}                     & \textbf{Stoichiometric vector}\\
\hline
$\mu \rightarrow Y_1$  & $a_1(x)   = \mu\Omega + o(\Delta t)$                  & $v_1   =[1,0,0,0,0]$\\
$Y_1 \rightarrow Y_2$  & $a_2(x)   = \beta_1 y_1y_2+ o(\Delta t)$        & $v_2   =[-1,1,0,1,0]$\\
$Y_1 \rightarrow Y_3$  & $a_3(x)   = \beta_2 y_1y_3+ o(\Delta t)$        & $v_3   =[-1,0,1,0,1]$\\
$Y_1 \rightarrow \mu$  & $a_4(x)   = \mu y_1+ o(\Delta t)$               & $v_4   =[-1,0,0,0,0]$\\
$Y_1 \rightarrow Y_4$  & $a_5(x)   = \upsilon y_1+ o(\Delta t)$          & $v_5   =[-1,0,0,1,0]$\\
$Y_2 \rightarrow Y_4$  & $a_6(x)   = \gamma y_2+ o(\Delta t)$            & $v_6   =[0,-1,0,0,0]$\\
$Y_2 \rightarrow \mu $ & $a_7(x)   = \mu y_2 + o(\Delta t)$              & $v_7    =[0,-1,0,0,0]$\\
$Y_5 \rightarrow Y_2$  & $a_8(x)   = \sigma\beta_1 y_5y_2+ o(\Delta t)$  & $v_8   =[0,1,0,0,-1]$\\
$Y_4 \rightarrow Y_3$  & $a_9(x)   = \sigma\beta_2 y_2y_4+ o(\Delta t)$  & $v_9   =[0,0,1,-1,0]$\\
$Y_3 \rightarrow Y_5$  & $a_{10}(x)= \gamma y_3+ o(\Delta t)$            & $v_{10}=[0,0,-1,0,0]$\\
$Y_3 \rightarrow \mu$  & $a_{11}(x)= \mu y_3+ o(\Delta t)$               & $v_{11}=[0,0,-1,0,0]$\\
$Y_4 \rightarrow \mu$  & $a_{12}(x)= \mu y_4+ o(\Delta t)$               & $v_{12}=[0,0,0,-1,0]$\\
$Y_5 \rightarrow \mu$  & $a_{13}(x)= \mu y_5+ o(\Delta t)$               & $v_{13}=[0,0,0,0,-1]$\\
\end{tabular}
\end{center}
\caption{\textit{List of reactions and stochiometric vectors for system (\ref{eq:ODE2}).}}
\label{tab:reactions}
\end{table}

The master equation is given by
\begin{equation}\label{eqs:PSD_ME}
P_y(t) = \sum_{i_1}^{\cal R}\{a_i(y-v_i)P_{y-v_i}(t)-a_i(y)P_y(t)\}
\end{equation}

The stochiometric vectors $v_i$ and the rate of reactions $a_i$,  $i=1,\ldots, \cal R$ are presented in Tab. \ref{tab:reactions}. The van Kampen expansion \cite{van1992stochastic} writes the number of individuals as a sum of two parts

\begin{equation}\label{eq:phi}
Y_k(t) = \Omega \phi_k(t)+ \Omega^{1/2}\xi_{k}(t)
\end{equation}

where $\phi_k(t)$, $k=1,\ldots,5$ describes the macroscopic behaviour. $\xi_k(t)$  represents the aggregate effects of demographic stochasticity and describes the fluctuations. We expand in power of $\Omega$ and collect powers of $\Omega^{1/2}$ to find the macroscopic law given by

\begin{align*}
\dfrac{d\phi_k}{dt} &= \sum_{i=1}^{\cal R} S_{kij}f'_i(\phi)\\
\dfrac{d\phi}{dt}   &= Sf(\phi)
\end{align*}

where  $S=[v_1,\ldots,v_{\cal R}]$ is the stoichiometric matrix and  $f(\phi)=[a_1(\phi), \ldots, a_{\cal R}(\phi)]$ is the vector with propensities. We set $\eta_1=\eta_2=0$. The expressions for the scaled macroscopic equations are:

\begin{equation}\label{eq:ODE}
\begin{split}
\dot{\phi_1}(t) &= -\beta_2\phi_3\phi_1-\beta_1\phi_2\phi_1-\upsilon\phi_1+\mu(1-\phi_1)\\
\dot{\phi_2}(t) &= \beta_1\phi_2(\phi_1+\sigma\phi_5)-(\gamma+\mu)\phi_2\\
\dot{\phi_3}(t) &= \beta_2\phi_3(\phi_1+\sigma\phi_4)-(\gamma+\mu)\phi_3\\
\dot{\phi_4}(t) &= \upsilon \phi_1 + \beta_1\phi_2\phi_1-\sigma\beta_2\phi_3\phi_4-\mu\phi_4\\
\dot{\phi_5}(t) &= \beta_2\phi_3\phi_1-\beta_1\sigma\phi_2\phi_5-\mu\phi_5\\
\end{split}
\end{equation}

To take into account the demographic stochasticity we will use the unscaled system:

\begin{equation}\label{eq:ODE3}
\begin{split}
\dot{\phi_1}(t) &= -\beta_2\dfrac{\phi_3}{\Omega}\phi_1-\beta_1\dfrac{\phi_2}{\Omega}\phi_1-\upsilon\phi_1+\mu(\Omega-\phi_1)\\
\dot{\phi_2}(t) &= \beta_1\dfrac{\phi_2}{\Omega}(\phi_1+\sigma\phi_5)-(\gamma+\mu)\phi_2\\
\dot{\phi_3}(t) &= \beta_2\dfrac{\phi_3}{\Omega}(\phi_1+\sigma\phi_4)-(\gamma+\mu)\phi_3\\
\dot{\phi_4}(t) &= \upsilon \phi_1 + \beta_1\dfrac{\phi_2}{\Omega}\phi_1-\sigma\beta_2\dfrac{\phi_3}{\Omega}\phi_4-\mu\phi_4\\
\dot{\phi_5}(t) &= \beta_2\dfrac{\phi_3}{\Omega}\phi_1-\beta_1\sigma\dfrac{\phi_2}{\Omega}\phi_5-\mu\phi_5\\
\end{split}
\end{equation}

Then, we collect powers of $\Omega^0$ to obtain a set of Langevin equations
\begin{equation}\label{eqs:PSD_langevin}
\dot{\xi}(t) = A(t)\xi(t)+\zeta(t)
\end{equation}

$\zeta(t)$ is the white noise with zero mean and its cross-correlation structure is given by $\langle\zeta(t)\zeta(t')^T\rangle = B(t)\delta(t-t')$. \\

\textbf{Matrices A and B}\\

Matrix $A(t)$ is represented by $A=S.\dfrac{\partial f(\phi)}{\partial \phi}$ which is equivalent to the Jacobian of system (\ref{eq:ODE3}); matrix $B(t)$ is given by $B= Sdiag(f(\phi))S^T$.\\ 


\begin{center}
\textbf{Matrix A}\\
\begin{small}
\resizebox{1 \textwidth}{!}{  
$\begin{pmatrix}\label{MatA}
-\beta_1\dfrac{\phi_2}{\Omega}-\beta_2\dfrac{\phi_3}{\Omega} -\upsilon-\mu & -\beta_1\dfrac{\phi_1}{\Omega} &-\beta_2\dfrac{\phi_1}{\Omega} & 0 & 0\\
\beta_1\dfrac{\phi_2}{\Omega} &\dfrac{\beta_1}{\Omega}(\phi_1+\phi_5\sigma) -\gamma-\mu&0&0& \beta_1\dfrac{\phi_2}{\Omega}\sigma\\
\beta_2\dfrac{\phi_3}{\Omega} & 0 &\dfrac{\beta_2}{\Omega}(\phi_1+\phi_4\sigma)-\gamma-\mu & \beta_2\dfrac{\phi_3}{\Omega}\sigma & 0\\
\beta_1\dfrac{\phi_2}{\Omega}+\upsilon & \beta_1\dfrac{\phi_1}{\Omega} &-\beta_2\dfrac{\phi_4}{\Omega}\sigma & -\beta_2\dfrac{\phi_3}{\Omega}\sigma-\mu & 0\\
\beta_2\dfrac{\phi_3}{\Omega} & -\beta_1\dfrac{\phi_5}{\Omega}\sigma &\beta_2\dfrac{\phi_1}{\Omega} &0&-\beta_1\dfrac{\phi_2}{\Omega}\sigma-\mu\\
\end{pmatrix}$}
\end{small}
\end{center}
\vspace{1cm}
\begin{center}
\textbf{Matrix B}\\
\begin{small}
\resizebox{1 \textwidth}{!}{  
$\begin{pmatrix}\label{MatB}
\beta_1\phi_1\dfrac{\phi_2}{\Omega} + \upsilon\phi_1 & -\beta_1\phi_1\dfrac{\phi_2}{\Omega} & -\beta_2\phi_1\dfrac{\phi_3}{\Omega} & -\beta_1\phi_1\dfrac{\phi_2}{\Omega}-\upsilon\phi_1 &-\beta_2\phi_1\dfrac{\phi_3}{\Omega}\\
+\beta_2\phi_1\dfrac{\phi_3}{\Omega}+\mu\phi_1+\mu &&&&\\
-\beta_1\phi_1\dfrac{\phi_2}{\Omega} & \beta_1\phi_1\dfrac{\phi_2}{\Omega}+\gamma\phi_2+ & 0& \beta_1\phi_1\dfrac{\phi_2}{\Omega} & -\beta_1\sigma\phi_2\dfrac{\phi_5}{\Omega} \\
&\beta_1\sigma\phi_2\dfrac{\phi_5}{\Omega}+\mu\phi_2 &&&\\
-\beta_2\phi_1\dfrac{\phi_3}{\Omega} & 0 & \beta_2\phi_1\dfrac{\phi_3}{\Omega}+\gamma\phi_3+\mu\phi_3+ & -\beta_2\sigma\phi_3\dfrac{\phi_4}{\Omega} & \beta_2\phi_1\dfrac{\phi_3}{\Omega} \\
&& \beta_2\sigma\phi_3\dfrac{\phi_4}{\Omega} &&\\
-\beta_1\phi_1\frac{\phi_2}{\Omega}-\upsilon\phi_1 & \beta_1\phi_1\dfrac{\phi_2}{\Omega} & -\beta_2\sigma\phi_3\dfrac{\phi_4}{\Omega} & \beta_1\phi_1\dfrac{\phi_2}{\Omega}+\upsilon\phi_1+ &0\\
&&& \beta_2\sigma\dfrac{\phi_3}{\Omega}\phi_4+\mu\phi_4 & \\
-\beta_2\phi_1\dfrac{\phi_3}{\Omega} & -\beta_1\sigma\phi_2\dfrac{\phi_5}{\Omega} & \beta_2\phi_1\dfrac{\phi_3}{\Omega} &0& \beta_2\phi_1\dfrac{\phi_3}{\Omega}+ \\
&&&& \beta_1\sigma\phi_2\dfrac{\phi_5}{\Omega}+\mu\phi_5 \\
\end{pmatrix}$}
\end{small}
\end{center}

\subsection{Equilibrium Points}
System (\ref{eq:ODE3}) has four fixed points, we can find only two analytically.\\

Let $\phi_1(t)=\phi_2(t)=0$. The disease-free equilibrium point is given by:
\begin{equation}\label{FixPoint1}
\phi_{10} = \Omega\left[\dfrac{\mu}{(\mu+\upsilon)},0,0,\dfrac{\upsilon}{(\mu+\upsilon)},0\right]
\end{equation}

Now,  let $\phi_3(t)=0$.  The equilibrium with people infected with influenza only is
\begin{equation}\label{FixPoint2}
\phi_{20} = \Omega\left[\dfrac{\gamma+\mu}{\beta_1},\dfrac{\mu}{\gamma+\mu}-\dfrac{\upsilon+\mu}{\beta_1},0,1-\dfrac{\gamma+\mu}{\beta_1},0\right]
\end{equation}

\subsection{The Reproduction Number:}
In the system (\ref{eq:ODE}), $\phi_1$ and $\phi_2$ correspond to the infection states. Matrix $F$ and $V$ in the scaled equilibrium point ($\phi_{10}/\Omega$) are given by \\

$F =\dfrac{1}{(\upsilon+\mu)} \begin{pmatrix}
\beta_1\mu & 0\\
0          &\dfrac{\beta_2(\mu+\sigma\upsilon)}{\mu+\upsilon}
\end{pmatrix}$;  $V = \begin{pmatrix}
(\gamma+\mu) & 0\\
0            & (\gamma+\mu)
\end{pmatrix}$\\

We refer the reader to Van den Driessche et al. \cite{van2002reproduction} for further details about $R_0$ calculation. The next generation matrix, $FV^{-1}$, has two eigenvalues given by:
$$R_1 = \dfrac{\mu\beta_1}{(\gamma+\mu)(\mu+\upsilon)}; \hspace{0.5cm} R_2 = \dfrac{\beta_2(\mu+\sigma\upsilon)}{(\gamma+\mu)(\mu+\upsilon)}$$

The two eigenvalues correspond to the reproduction numbers for each pathogen, $R_1$ for influenza and $R_2$ for RSV. The maximum of the two is the basic reproduction number for the system is. Thus
\begin{equation}
R_0 = \max_{p=1,2} R_p
\end{equation}

The equilibrium point $\phi_{10}$ is stable if $R_0<1$ \cite{van2002reproduction}-, this implies that $R_1<1$ and $R_2<1$.

\section{Power Spectral Density Computation}

\subsection{Unforced Model}

\begin{algorithm}[H]
\caption{PSD - Unforced Model}
\DontPrintSemicolon
\KwData{Parameters values $p$,  discretization of frecuency ($\omega$)}
\KwResult{$PSD_k$ array ( $dim(PSD_k)=(1\times len(Y))$}
\Begin{
Find the stable equilibrium point of system (\ref{eq:ODE3}) ($\phi_0$)\;
Evaluate matrix A and B in $p$ and $\phi_0$\;
\For{$j$ in $\omega$}{
      Compute matrix $\Phi(\omega_j) = -i\omega_j I-A$\;
      Compute matrix $P(\omega_j) = \Phi(\omega)B \Phi^\dagger(\omega)$\;
      Save   $PSD_k(\omega_j) = Re(P_{kk}(\omega_j))$ for $k=1,\ldots,dim(\phi)$\;
     }}
\end{algorithm}

\bibliography{Bibliography}{}
\bibliographystyle{plain}